\def\tech{0}
\def\fb{0}

\ifnum\fb=1
\documentclass[10pt, sigconf, screen]{acmart}
\else
\documentclass[letterpaper,twocolumn,10pt]{article}
\usepackage{usenix}
\fi

\usepackage{fancyhdr}
\usepackage{epsfig,endnotes}
\usepackage{epstopdf}
\usepackage{times}
\usepackage[subtle,tracking=normal,margins=normal,leading=normal,title=normal]{savetrees}
\usepackage{comment}
\usepackage{graphicx}
\usepackage{xspace}
\usepackage{pslatex}
\usepackage{amsmath,amssymb,empheq}
\usepackage{upgreek}
\usepackage[boxed]{algorithm2e}
\usepackage{multirow}
\usepackage{tabularx}
\DeclareGraphicsExtensions{.eps,.jpg,.png,.gif}
\graphicspath{{./figures/}}
\usepackage{color}
\usepackage{ifpdf}
\usepackage{mathptmx}  %
\usepackage{xfrac}
\usepackage[normalem]{ulem}  %
\usepackage{microtype} %
\usepackage{mathtools}

\ifnum\fb=1
\PassOptionsToPackage{hyphens}{url}
\else
\usepackage[hyphens]{url}
\fi

\usepackage[font=small,aboveskip=6pt,belowskip=-6mm]{caption}
\DeclareCaptionType[placement=b,within=none]{copyrightbox}
\usepackage{subfig}

\newcommand{\figref}[1]{Fig.~\ref{fig:#1}}
\newcommand{\secref}[1]{Sec.~\ref{sec:#1}}
\newcommand{\tabref}[1]{Tab.~\ref{tab:#1}}

\newcommand{\tr}[1]{#1}
\newcommand{\ntr}[1]{#1}
\ifnum\tech=1
\renewcommand{\ntr}[1]{}
\else
\renewcommand{\tr}[1]{}
\fi

\DeclareMathOperator*{\argmin}{arg\,min}

\makeatletter
  \newcommand\figcaption{\def\@captype{figure}\caption}
  \newcommand\tabcaption{\def\@captype{table}\caption}
\makeatother

\newcommand{\name}[0]{Volur\xspace}
\newcommand{\nameFL}[0]{Volur-FL\xspace}
\newcommand{\nameLB}[0]{Volur-LB\xspace}

\usepackage{booktabs}

\usepackage{enumitem}

\newenvironment{vinlist}
{\vspace{-4pt}\begin{itemize}[leftmargin=1.25em]
  \setlength{\itemsep}{-1pt}
  \setlength{\labelwidth}{1em}
  \setlength{\labelsep}{0.5em}
  \setlength{\topsep}{0pt}
  \setlength{\partopsep}{0pt}}
{\end{itemize}\vspace{-4pt}}

\newenvironment{vinenum}
{\vspace{-4pt}\begin{enumerate}[leftmargin=1.5em]
  \setlength{\itemsep}{-1pt}
  \setlength{\labelwidth}{1em}
  \setlength{\labelsep}{0.5em}
  \setlength{\topsep}{0pt}
  \setlength{\partopsep}{0pt}}
{\end{enumerate}\vspace{-4pt}}

\newcommand{\heading}[1]{\vspace{4pt}\noindent\textbf{#1}}
\newcommand{\topheading}[1]{\noindent\textbf{#1}}

\ifpdf
\else
\fi

\newcommand*{\affaddr}[1]{#1} %
\newcommand*{\affmark}[1][*]{\textsuperscript{#1}}

\title{\name: Concurrent Edge/Core Route Control in Data Center Networks}

\author{
       {\normalsize {\rm Qiao Zhang}\affmark[*]},
       {\normalsize {\rm Danyang Zhuo}\affmark[*]},
       {\normalsize {\rm Vincent Liu}\affmark[$\dagger$]},
       {\normalsize {\rm Petr Lapukhov}\affmark[$\ddagger$]},
       {\normalsize {\rm Simon Peter}\affmark[$\S$]},
       {\normalsize {\rm Arvind Krishnamurthy}\affmark[*]},
       {\normalsize {\rm Thomas Anderson}\affmark[*]}\\
\affaddr{\affmark[*]University of Washington, \affmark[$\dagger$]University of Pennsylvania, \affmark[$\ddagger$]Facebook, \affmark[$\S$]University of Texas at Austin}\\
}

\date{}
\begin{document}

\pagestyle{plain}

\maketitle

\newcommand{\tts}[1]{\texttt{\small #1}}

\definecolor{Orange}{rgb}{1,0.5,0}
\definecolor{Gray}{rgb}{0.5,0.5,0.5}
\definecolor{LtGray}{rgb}{0.66,0.66,0.66}
\definecolor{Red}{rgb}{0.8,0,0}
\definecolor{Green}{rgb}{0,0.33,0}

\newcommand{\brok}[1]{\sout{#1}}
\newcommand{\cbrok}[1]{\textcolor{Gray}{\brok{#1}}}
\newcommand{\less}[1]{\ensuremath{\check{#1}}}
\newcommand{\cless}[1]{\textcolor{Gray}{\less{#1}}}
\newcommand{\ok}[1]{#1}
\newcommand{\cok}[1]{\ok{#1}}

\renewcommand{\b}[1]{\cbrok{#1}}
\renewcommand{\l}[1]{\cless{#1}}
\renewcommand{\o}[1]{\cok{#1}}

\newcommand{\vhead}[1]{\begin{sideways}#1\end{sideways}}
\newcommand{\none}{-}
\newcommand{\broken}{\ensuremath{\otimes}}
\newcommand{\lessfunc}{\ensuremath{\circleddash}}
\newcommand{\phone}{Phone \#}
\newcommand{\histbookmarks}{Bookmarks}
\newcommand{\sms}{SMS}
\newcommand{\imei}{Device ID}
\newcommand{\location}{Location}
\newcommand{\contacts}{Contacts}
\newcommand{\calendar}{Calendar}
\newcommand{\verify}[1]{\textcolor{Red}{#1}}

\newcommand{\sur}[0]{S}
\newcommand{\cov}[0]{C}
\newcommand{\ove}[0]{O}

\newcommand{\ooo}[0]{\o{\sur}\o{\cov}\o{\ove}}
\newcommand{\ool}[0]{\o{\sur}\o{\cov}\l{\ove}}
\newcommand{\oob}[0]{\o{\sur}\o{\cov}\b{\ove}}
\newcommand{\olo}[0]{\o{\sur}\l{\cov}\o{\ove}}
\newcommand{\oll}[0]{\o{\sur}\l{\cov}\l{\ove}}
\newcommand{\olb}[0]{\o{\sur}\l{\cov}\b{\ove}}
\newcommand{\obo}[0]{\o{\sur}\b{\cov}\o{\ove}}
\newcommand{\obl}[0]{\o{\sur}\b{\cov}\l{\ove}}
\newcommand{\obb}[0]{\o{\sur}\b{\cov}\b{\ove}}

\newcommand{\loo}[0]{\l{\sur}\o{\cov}\o{\ove}}
\newcommand{\lol}[0]{\l{\sur}\o{\cov}\l{\ove}}
\newcommand{\lob}[0]{\l{\sur}\o{\cov}\b{\ove}}
\newcommand{\llo}[0]{\l{\sur}\l{\cov}\o{\ove}}
\newcommand{\llll}[0]{\l{\sur}\l{\cov}\l{\ove}} %
\newcommand{\llb}[0]{\l{\sur}\l{\cov}\b{\ove}} %
\newcommand{\lbo}[0]{\l{\sur}\b{\cov}\o{\ove}}
\newcommand{\lbl}[0]{\l{\sur}\b{\cov}\l{\ove}} %
\newcommand{\lbb}[0]{\l{\sur}\b{\cov}\b{\ove}} %

\newcommand{\boo}[0]{\b{\sur}\o{\cov}\o{\ove}}
\newcommand{\bol}[0]{\b{\sur}\o{\cov}\l{\ove}}
\newcommand{\bob}[0]{\b{\sur}\o{\cov}\b{\ove}}
\newcommand{\blo}[0]{\b{\sur}\l{\cov}\o{\ove}}
\newcommand{\bll}[0]{\b{\sur}\l{\cov}\l{\ove}}  %
\newcommand{\blb}[0]{\b{\sur}\l{\cov}\b{\ove}}  %
\newcommand{\bbo}[0]{\b{\sur}\b{\cov}\o{\ove}}
\newcommand{\bbl}[0]{\b{\sur}\b{\cov}\l{\ove}}  %
\newcommand{\bbb}[0]{\textcolor{red}{\brok{\sur}\brok{\cov}\brok{\ove}}}

\newcommand{\vladd}[1]{\textcolor{cyan}{#1}}
\newcommand{\vlremove}[1]{\textcolor{Gray}{\sout{#1}}}
\newcommand{\vlreplace}[2]{\textcolor{Gray}{\sout{#1}}\textcolor{cyan}{#2}}
\newcommand{\vlcomment}[1]{\textcolor{Orange}{[vl: #1]}}

\newcommand{\qzadd}[1]{\textcolor{magenta}{#1}}
\newcommand{\qzremove}[1]{\textcolor{Gray}{\sout{#1}}}
\newcommand{\qzreplace}[2]{\textcolor{Gray}{\sout{#1}}\textcolor{magenta}{#2}}
\newcommand{\qzcomment}[1]{\textcolor{Orange}{[qz: #1]}}

\newcommand{\spadd}[1]{\textcolor{violet}{#1}}
\newcommand{\spremove}[1]{\textcolor{Gray}{\sout{#1}}}
\newcommand{\spreplace}[2]{\textcolor{Gray}{\sout{#1}}\textcolor{violet}{#2}}
\newcommand{\spcomment}[1]{\textcolor{Orange}{[sp: #1]}}

\subsection*{Abstract}

A perennial question in computer networks is where to place functionality among components of a distributed computer system.
In data centers, one option is to move all intelligence to the edge, essentially relegating switches and middleboxes, regardless of their programmability, to simple static routing policies.
Another is to add more intelligence to the middle of the network in the hopes that it can handle any issue that arises.

This paper presents an architecture, called \name, that provides a third option by facilitating the co-existence of an intelligent network with an intelligent edge.
The key architectural principle of \name is \textit{predictability} of the network.
We describe the key design requirements, and show through case studies how our approach facilitates more democratic innovation of all parts of the network.
We also demonstrate the practicality of our architecture by describing how to implement the architecture on top of existing hardware and by deploying a prototype on top of a large production data center.

\vspace{-1ex}
\section{Introduction}

A perennial question in computer networks is where to place routing functionality among components of a distributed computer system: whether it be at the end hosts or in the network itself.
In data centers in particular, the research community has presented compelling arguments for route control and visibility at both locations.
Their work has shown the importance of placement to critical network properties like fault tolerance and load balancing.

Link/switch failure handling is a good example of the complexities of this decision.
One broad class of proposals argues for giving the network the ability to detect and route around failures~\cite{Everflow,FlowRadar,F10}---in essence, they argue for a \textit{smart} network supporting a \textit{simple} edge.
When working as intended, these systems are both fast and efficient at handling failures; however, for a broad class of failures (e.g., ``silent" failures), the detection methods themselves can fail, leaving end hosts with little-to-no visibility or control of how their packets are handled in the network.

The other class argues for moving all failure handling to the edge of the network, essentially relegating switches and middleboxes to simple static routing policies~\cite{Openflow}, i.e., a \textit{simple} network controlled by a \textit{smart} edge.
In this approach, fate sharing guarantees that no packet losses will go unnoticed; however, this comes at the cost of the ability to react to easily detectible failures quickly and locally---a feature that is essential to high network availability.

This paper explores a third option: the co-existence of an intelligent network with an intelligent edge.
Even constraining ourselves to fault tolerance and load balancing, certain problems within those domains are best implemented in the former, while others are best implemented at the latter---an ideal network architecture would allow for both.
To that end, we present a new data center network architecture, \name, that facilitates this interaction.
In the end, however, is not possible to allow all features at all locations in the network (fine-grained load balancing, for instance, relies on transient information that is typically not externally visible).
Instead, the goal of \name is to provide clear guidelines for where to implement features, to detail the restrictions on those features, and to present a framework to implement them in a conforming way.

The key architectural requirement of \name is \textit{predictability of the network}.
Specifically, that switch routing behavior should be externally predictable by the trusted network-layer software running on the endpoints.
Predictability forms the contract between the network and its end hosts; as long as routing decisions are predictable and/or infrequent, switches are free to do as they wish.
In return, end hosts must allow for transient inaccuracies in prediction.
As an example, switches are allowed to locally detect and reroute around failures as long as end hosts eventually become informed of the new network state.

Our system, \name, presents a prototype implementation of predictable networking that is composed of three components: (1) switches that route using predictable functions of the packet's header and switch
configuration state, (2) a network state service that disseminates any required information to end hosts, and (3) a per-end host path choice module that models network behavior.
To demonstrate its flexibility, we implement two end host applications on top of \name that utilize the predictability of the network to locate failures and balance load in spite of concurrent in-network functionality.

To demonstrate its feasibility in practice, we built \name network predictors for two different deployments: a large production data center at Facebook with upwards of one hundred thousand devices~\cite{fb_datacenter}, and a smaller testbed.
These deployments span multiple switch ASICs in switches from multiple vendors, and show that while switch functionality \textit{can} be complicated, it does not \textit{need} to be complicated.
While we understand that not every network operator has the necessary information to implement this approach today, it is our hope that the benefits we show provide sufficient incentives for future development in this direction.

Finally, to evaluate our architecture, we utilized the aforementioned large production data center
deployment; a second, modestly-sized Cloudlab testbed; and an ns-3 packet-level
simulated network.
Using these testing environments, we show: (1) that our predictability-supported end host failure handling system can locate non-fail-stop failures with over 0.95 precision and 0.85 recall even if there are multiple, diverse failures and even if not all hosts participate in localization, and (2) that hosts can route around those failures within a fraction of a second.
We also show for at least one network feature that is disallowed in our architecture---fine-grained load balancing---end-host-based design on our architecture approach state-of-the-art in-network approaches like CONGA.

More specifically, we make the following contributions:

\begin{vinlist}
\item We present the design and implementation of a novel data center architecture, \name, that facilitates the co-existence of an intelligent network with an intelligent edge through the contract of predictability.
\item We introduce a system, \nameFL, that leverages predictability to implement extremely accurate and fine-grained failure localization and rerouting in the presence of relatively-complex network features.
\item We also introduce a second system, \nameLB, that demonstrates both the flexibility of \name and how to emulate in-network features predictably.
\item Finally, we demonstrate the practicality of \name by implementing a prototype that can accurately predict and control path choice on unmodified switches in a large-scale production data center.
\end{vinlist}
\vspace{-1em}
 
\section{Motivation}

\begin{figure}[t]
\centering
\vspace{-0.5ex}
\epsfig{file=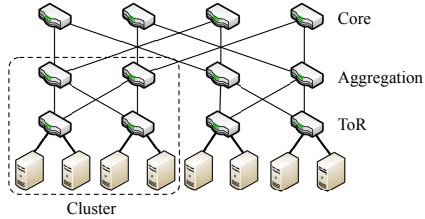, width=.7\columnwidth}
\vspace{-0.5ex}
\caption{A canonical 3-level Folded-Clos topology.  The three levels include ToR, aggregation, and core switches.  The high path diversity of these networks has contributed to an array of complex and opaque routing strategies.}
\label{fig:fattree}
\end{figure}

Today's data center networks typically take the form of a multi-rooted tree of switches like the one shown in \figref{fattree}.
One natural property of these tree topologies is the presence of many paths between any two end hosts.
Routing protocols, which select the path to use for any particular packet, are essential to both maintaining network reliability and ensuring balanced load.
These protocols are often complex and opaque to end hosts.

Central to this ecosystem is ECMP, a switch-level mechanism that randomly chooses among several options a next hop for each flow.
Other examples include Link Aggregation Groups (LAGs), which operate similarly
to ECMP, but among point-to-point links rather than paths;
Resilient Hashing~\cite{CumulusResilient}, which dictates ECMP behavior
so that flow assignment is stable even as links are brought up or taken down;
and more recently, in-network load balancing like LocalFlow~\cite{LocalFlow}, DRILL~\cite{drill}, DLB~\cite{DLB},
and CONGA~\cite{conga}, which route based on transient workload statistics.
Recent proposals for increased programmability of networks~\cite{P4} only increase the potential for complexity and opacity of routing.

\subsection{The Case for End Host Control}
\label{sec:case_endhost}

In contrast to the current state of the network, the research community has, over the years, made many compelling arguments for end host visibility and/or control of routing.
Some have noted that, for some features, end hosts are uniquely suited to solving a particular problem such as failure localization~\cite{XPATH} or performance isolation~\cite{XPATH}.
Others have noted that end host changes are easier to implement and deploy compared to changes to the network~\cite{flowbender, LetFlow}.

A classic example (and part of the original inspiration for the end to end argument~\cite{Saltzer1981EndToEndAI}) is the handling of packet drops and their associated network failures.
The opacity of today's routing protocols presents a significant challenge to the identification and mitigation of these failures, particularly when they evade traditional debugging tools such as heartbeats (e.g., BFD~\cite{rfc5880} and BGP keepalives) and switch drop counters.
Examples of failures that are not easily caught by traditional network features are:

\begin{vinlist}
\item \textit{Partial failures:}
Some failures are stochastic in nature~\cite{Everflow}.
Switches may not detect these as their occasional heartbeats/keepalives can get (un)lucky and miss the problem while application traffic will continue to experience packet drops.
\item \textit{Input-dependent failures:}
Some are not random, but instead only affect particular flows~\cite{Everflow, NetNORAD}.
Here too, heartbeats may not trigger the same routing path as the problematic traffic, and the switch may not notice the error.
\item \textit{Silent failures:}
Finally, switch counters are sometimes unreliable, leading to cases of silent failures that are not reflected in any network statistic.
For failures that are not noticed by the switch itself (e.g., partial or input-dependent) and are also not reflected in counters, detection is extremely difficult.
These are known to occur and cause significant headache in practice~\cite{AWSgray,Everflow,NetNORAD}.
\end{vinlist}

It is because of the above classes of failures that end hosts are often seen as an attractive location in which to detect and handle failures.
Researchers have also presented many other use cases for visibility and control of routing at the end host.
Unfortunately, the opacity of today's data center networks prevents this, limiting the scope/deployment of such approaches.

\subsection{The Case for Network Control}

A natural reaction to the desire for end host control over routing is to migrate the complexity of the network to the edge.
This approach is explored by several recent data center networking proposals~\cite{XPATH,CLOVE,LetFlow,roy17passive}.
Some of these proposals let networks handle routing, but give end hosts the ability to change paths on demand through an IPv6 Flow Label or similar mechanism~\cite{LetFlow,ResilientLB}---a useful workaround, but limited in the features it can support.

On the more extreme end of this spectrum are source routing approaches like those proposed in \cite{XPATH} or \cite{MPLS}.
These proposals successfully enable end hosts to perform fine-grained failure detection and rerouting, but a naive application of source routing to data centers surrenders at least two crucial features:

\begin{vinlist}
\item \textit{Fast failover:} Easily-detectable failures like signal loss on a link are more simply and quickly handled in the network.
In these cases, switches adjacent to the failure are able to detect/reroute at timescales orders of magnitude faster than the edge.
This fast failover is essential for achieving high network availability.
\item \textit{Backward compatibility:} Most current applications and operating systems are designed to be agnostic to the routing decisions of the network.
While it is possible to change all applications, OSes, and/or hypervisors, an ideal solution would permit the use of legacy software.
\end{vinlist}
\vspace{-1em}
 
\section{\name: A Predictable Network}

\begin{table*}[t]
\centering
\footnotesize
\begin{tabularx}{\textwidth}{|X|X|l|}
\hline
\textbf{Challenges} & \textbf{Solution} & \textbf{Section(s)}\\ \hline
Is it possible and/or practical to predict network behavior?%
		& Some networks are already predictable.  More generally, we anticipate that the OpenFlow model may also apply here---if customers value predictability, vendors will provide it as a feature.%
		& \ref{sec:design_switch}, \ref{sec:prototype}\\ \hline
\multirow{2}{*}{\shortstack[l]{Is predictability compatible with dynamic switch behavior?}}%
		& For infrequently-changing behavior (e.g., failover),  \name disseminates versioned network state to end hosts.%
		& \ref{sec:topo_service} \\ \cline{2-3}
		& For frequently-changing behavior (e.g., load balancing), end hosts can approximate current switch features.%
		& \ref{sec:loadbalancing} \\ \hline
How do we deal with unpredictable failures and other inaccuracies in prediction?%
		& End-hosts use versioned topology to sieve out reliable drop statistics. Common-case consistent hashing limit routing changes. %
		& \ref{sec:collection}, \ref{sec:eval_mispredicts} \\ \hline
How do we defend against DDoS attacks that might be enabled by end host path prediction/control?%
		& Only convey topology to the end host trusted computing base. A NAT can be used if extra protection is needed.%
		& \ref{sec:endhosts} \\ \hline
\end{tabularx}
\caption{A roadmap of key challenges in creating a predictable network and their solutions.}
\label{tab:issues_table}
\end{table*}

In this paper, we explore the design of an architecture for the peaceful co-existence of an intelligent network with intelligent end hosts.
The key architectural principle of our work is \textit{predictability of the network} as a method for interoperation.
In this model, switches are free to implement a wide range of routing techniques as long as they are externally predictable.
End hosts are then free to implement any functionality they wish on top of the predictable network.

More specifically, our requirement is that switches route based only on the packet header and infrequently changing configuration state.
Compared to pure source routing, prediction in this model is not always accurate.
The point of fast failover, for instance, is that the switches know about and can react to failures faster than end hosts.
Immediately after a failure, the network may not operate as end hosts expect.
Instead, end hosts must tolerate a small amount of inaccuracy in return for these features.

\begin{figure}[t]
\vspace{-10px}
\centering
\epsfig{file=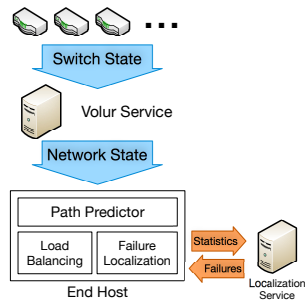, width=.5\columnwidth}
\caption{The primary components in \name, our instantiation of a predictable network.  A logically centralized \name Service collects the current configuration of the network and disseminates it to end hosts, which use it to predict paths.  Predictions can be used for many purposes, from locating failures to balancing load.}
\label{fig:overview}
\end{figure}

There are several challenges in making such a system practical, which we list in \tabref{issues_table}.
Thus, the primary contribution of our work is to characterize what it takes to design and implement a predictable network and to detail its benefits/limitations.
\name consists of three primary components: \textit{switches} that are predictable,
a \textit{\name service} that gathers and distributes the current state 
of switches, and \textit{end hosts} that use that state to predict routes.
The overall architecture is illustrated in \figref{overview}.

\subsection{A Predictable Switch}
\label{sec:design_switch}

As mentioned, our primary design principle is simple to state: switches should route only on the packet's header and infrequently changing configuration state.
Note that this restriction only applies to functions that affect the packet's path---features such as management, monitoring, QoS, and queuing are all orthogonal.

\subsubsection{A Simple Predictable Packet Pipeline}

To see why our design principle is congruent with common-case network features, we describe the implementation of a predictable network router.
Due to space constraints, we limit our exposition to the subset of the pipeline necessary for forwarding a simple Ethernet and IPv4 packet without VLANs or tunneling.
The switch we describe allows for both fast failover and backward compatibility.

In \secref{prototype}, we go on to show that, not only can we build a predictable switch, we can also configure some existing switches to be predictable.

\heading{L2 processing.}
L2 processing is typically based purely on table lookups.
For instance, if the destination MAC of the packet matches the switch's MAC address, the packet will continue to L3 processing.
Otherwise, it will be switched as a raw Ethernet frame (we omit those details).

\noindent
\textit{Depends on: packet header and switch's MAC address.}

\heading{L3 processing.}
L3 processing is also based on table lookups, but may require other features as well.
The switch begins by looking up the destination IP in its forwarding table.
The resulting entry may either point to an egress port, multiple egress ports, or indicate that the packet should be dropped.
When there are multiple possible next hops, as is often the case in Clos networks, the switch will
calculate a hash function over several subfields of each packet's header.
The result of the hash function (modulo the number of possible next hops) is used as an index into the next hop table to determine the egress port.

ECMP has traditionally been considered unpredictable, but for efficiency, modern ECMP implementations are typically deterministic.
For instance, it might hash over a packet's 5-tuple using simple functions like XOR, CRC, or table lookups~\cite{kalkunte2010high,rfc2992,davies2010traffic}.
In practice, these hash functions can be combined with hash seeds, preprocessing, bit shifting, masks, and resilient hashing techniques to improve results in various situations~\cite{CumulusResilient}, but
all of the above functions are predictable as long as changes are relatively infrequent.

\noindent
\textit{Depends on: packet header, L3 forwarding table, multipath table, and ECMP hash configuration.}

\heading{Egress modifications.}
Finally, before the packet is sent back out on the wire, the switch will update the src and dst MAC addresses to correspond to the next L2 hop.
In addition, it will recalculate the TTL field and IP and Ethernet checksums.

\noindent
\textit{Depends on: switch's MAC address, neighbor's MAC address, and packet header.}

\subsubsection{Other Network Routing Functions}
\label{sec:design_other}

The above discussion focuses on L3 forwarding in a predictable switch: how to implement it and how to predict it given the current state of the switch.
ECMP is included in the set of functions that can be made predictable in this fashion.
The same is true of most other forwarding functionality, e.g., encapsulation, VLANs, and QoS.
There are, however, some switch routing features that cannot be made predictable.
These typically involve fine-grained load balancing, e.g., CONGA~\cite{conga} and DRILL~\cite{drill}. 
We can classify functions into these two categories based on their inputs:

\heading{Infrequently changing state:}
If, in addition to the packet header, the switch routing algorithm depends only on infrequently-changing  state, it is considered to be predictable in our model.
As an example, in L3 routing, failures and routing updates can cause unpredictable changes in the network, but as long as the changes are infrequent, end hosts network applications are expected to handle those inaccuracies.
The aforementioned forwarding functionality falls squarely into this category, as do many recent data center routing proposals including WCMP~\cite{WCMP}, F10~\cite{F10}, B4~\cite{B4}, and SWAN~\cite{SWAN}

\heading{Frequently changing state:}
If, on the other hand, the switch routing algorithm depends on frequently changing state such as instantaneous queue length or utilization, the function is considered unpredictable.
The cutoff for frequency is determined by the operator and is a function of the accuracy requirements of route prediction.
Examples of algorithms in this category include DLB~\cite{DLB}, CONGA~\cite{conga}, DRILL~\cite{drill}, and LocalFlow~\cite{LocalFlow}---all proposals for fine-grained, in-network load balancing.
They also include certain counter-based ACL, QoS, and packet processing policies found on modern switches.
These functions are disallowed in \name switches, but in \secref{loadbalancing} we explore the efficient and accurate end host emulation of this class of proposals.

\subsection{The \name Service}
\label{sec:topo_service}

Predicting the route of a packet requires both the packet's header and elements of the switch's current state.
For the sender of the packet, obtaining its header is simple.
For the other piece of information---switch state---we introduce an aggregation service that gathers up-to-date state from every switch and disseminates it to every end host.
This dissemination must be performed on any switch state change including link failures/recoveries and control plane routing updates.
Replication and sharding of such a service is straightforward; for ease of explanation, we assume a logically centralized \name service. %

The primary goal of the \name service is to disseminate switch updates as quickly as possible.
There are two steps:

\heading{Switches to the \name service.}
As state updates may occur at irregular intervals and must be disseminated quickly, switches mostly operate on a push model.
When a state change occurs (e.g., a BGP update or link failure), switches will immediately send a \texttt{diff} of their state to the \name service.
The service also periodically pings each switch for a hash of their current state to ensure that it is still alive and correctly synchronized.
In systems with an existing centralized SDN infrastructure, the \name service is a natural extension to the SDN controller.

\begin{figure}
\centering
\epsfig{file=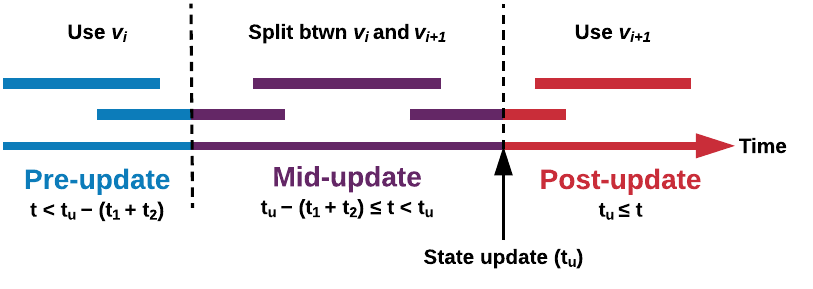, width=0.9\columnwidth}
\caption{End hosts classify traffic into three classes given a state update at time $t_u$: pre-update, mid-update, and post-update.}
\vspace{-1px}
\label{fig:topo_change_timeline}
\end{figure}

\heading{\name service to end hosts.}
The second step is to disseminate the state changes to end hosts.
There are two channels for state dissemination in our system.
The end hosts periodically pull a full snapshot of the current network state.
In addition, the \name service broadcasts versioned, perishable state updates to end hosts.
These updates are sent using UDP to ensure time bounds.

\begin{vinenum}
\item When a switch updates its state, it sends a \texttt{diff} of the state change to the \name service.  Let the maximum propagation delay of this message be $t_1$.
\item Upon receipt of the update, \name increments its version number $v\rightarrow v+1$ and distributes the update to all affected hosts.  Let the maximum delay of this message be $t_2$.
\item Upon receipt of an update from the \name service, hosts send back an acknowledgment.
\item If the ACK is not received after some predetermined timeout, inform the end host during its next checkpoint.
\end{vinenum}

Given the above protocol, we can quantify the length of three distinct phases of prediction accuracy.
Assume that the end host receives an update at time $t$, as shown in \figref{topo_change_timeline}.
\textit{Pre-change} predictions (before $t-(t_1+t_2)$) are correct.
\textit{Mid-update} predictions are slightly uncertain in that they can follow either state $v$ or $v+1$.  This period lasts from $t-(t_1+t_2)$ to $t$.
\textit{Post-update} predictions starting from time $t$ should all follow version $v+1$.
If the end host, during a checkpoint, finds out that it failed to acknowledge an update, all predictions between the current checkpoint and the last one are potentially inaccurate.
All inaccuracies are handled by higher-level network applications.

\subsection{End Hosts}
\label{sec:endhosts}

Given a predictable network, end host operation is relatively straightforward.
We provide to each end host a switch predictor that takes a switch state object and an input packet header.
The output of the predictor is a next hop and output packet header.

\heading{Predicting a packet's path.}
For every packet, path prediction is just a matter of iteratively chaining the next hop and packet header predictions of each intervening switch.
The switch state object is obtained from the \name service as described above; the end hosts already have the initial packet header.

\heading{Controlling a packet's path.}
To control a packet's path, end hosts only need to find a packet header that maps onto a target/acceptable round-trip path.
As switch operations are typically not cryptographically secure, it is often possible to create an efficient inverse for them.
In \secref{prototype}, we show that such techniques can be used to generate headers for specific paths in our large production data center network in under 12\,$\upmu$s.
Solutions are not guaranteed to exist, but operators work hard to ensure that hash functions cover the entire network evenly.%

End hosts have at least a few options when trying to craft a header to hit one of $n$ paths.
In general, they need $\sim\lceil\log n\rceil$ bits that are otherwise unused by the network.
For instance: \textit{IPv6 flow labels (20 bits)} are intended for purposes like ours; \textit{port numbers (up to 32 bits)} can also be used, but in the case of source ports, this may require minor OS changes in the way ports are allocated; and finally, \textit{IP addresses (up to 64/256 bits)} are possible as well, for instance by giving each server a /24 or /120.
End hosts can also combine bit regions to obtain a larger `address space'.

Legacy hosts can continue to send packets without modification, and those packets will be load balanced with ECMP just as they are today.

\heading{Preventing malicious control of paths.}
As a corollary, our architecture allows for efficient defenses against DDoS attacks.
A potential concern with our system is that it may allow malicious users in multi-tenant data centers to launch a targeted DDoS attack against individual network elements.
To that end, we note that without up-to-date switch state, the network is not more predictable than it is today---the configuration state space is very large and constantly changing.
Further, because cluster and fabric switches and links have extremely high capacity, it would be difficult for the attacker to determine whether any particular trial succeeded at steering to a particular path without access to data center internal traceroute.
Thus, the \name service only distributes state to the trusted computing base, and not untrusted applications/VMs.

Even so, if more security is necessary, the VM layer can pick a random source port or flow label for each connection, similar to the NATs that many VMMs already use.
If even a small part of the header is randomized, steering is difficult.

\heading{Changing paths mid-connection.}
\label{sec:flow_reroute_rules}
Beyond controlling a single packet's path is controlling the path of an entire TCP connection.
For new connections, this is just a matter of choosing a suitable 5-tuple for the connection.
To reroute existing connections
to avoid a failed or congested network component,
\name must change the packet headers without disrupting TCP's ability to demultiplex traffic.
IPv6 flow labels are a good candidate for this.
Otherwise, e.g., in the case of TCP source ports, the VMM/OS may need to rewrite the packet headers.

To be more concrete about this second option, when Alice wishes to change the path of a connection with Bob, she might decide on a TCP source port, $s$, that results in the target forward and reverse paths.
Alice will send the new source port to Bob asynchronously in a separate connection.
This must be done out-of-band because in the case of a failed path, the original connection may not be usable.
When Bob acknowledges the new source port, Alice installs packet rewrite rules as follows:

\begin{vinlist}
\item \textit{Egress:} Alice overwrites the src port number with
$s$.
\item \textit{Ingress:} Alice remembers the original src port number in a hash table so that when a response comes in, she can insert the original port transparently.
\end{vinlist}

\noindent
Bob installs similar rewrite rules:

\begin{vinlist}
\item \textit{Egress:} Bob overwrites the dst port number with $s$.
\item \textit{Ingress:} Bob remembers the original dst port number in a hash table so that when a response comes in, he can insert the original port transparently.
\end{vinlist}

Both ends of the connection can initiate such a path change, but to prevent flapping, we designate the client that called \texttt{connect()} to be responsible for most path changes.
\vspace{-1em}
\section{Case Studies}
\vspace{-1em}

Predictability provides a rich interface for end hosts, and we show two uses of that predictability.
The first is a fault localization service that showcases the flexibility of end hosts in \name despite static load balancing and failure reaction in the network.
The second is a load balancing mechanism that simultaneously demonstrates the power of our approach and shows how to emulate state-of-the-art dynamic network behavior predictably.

\vspace{-1em}
\subsection{\nameFL: Fault Localization}
\label{sec:localization}

The goal of \nameFL is attribute packet drops to specific components.
At a high level, we model the fault localization problem as an optimization problem~\cite{lgorzata2004survey,Gestalt}.
The intuition is that if many lossy paths from many vantage points across the data center converge at a single component, we can implicate the component as possibly faulty.
We show how careful accounting and analysis can overcome any inaccuracies that may arise from concurrent network changes.

\subsubsection{Collecting Drop Statistics}
\label{sec:collection}

\nameFL first collects drop statistics for each path.
For TCP traffic (the majority of data center traffic), drop information is already readily available in the form of retransmission statistics.
\nameFL uses Linux eBPF (Extended Berkeley Packet Filters)~\cite{eBPF} to gather these statistics on a per-connection basis.

Specifically, we track two TCP variables: \textit{pktsSent}, the number of packets sent and \textit{pktsRetrans}, the number of packets retransmitted.
Hosts poll these statistics every 10\,s.
Note that these variables track control packets that are ACKed (e.g., SYN/FIN packets) in addition to data packets.
It is important to track control packets since, for black holes, no data packets will be sent, only SYNs.
These statistics are approximations of the ground truth as in-flight packets, cumulative ACKs, and spurious retransmits can affect these numbers; however, our evaluations show that this approximation is effective in practice.

Non-TCP traffic is slightly more complex as not all protocols acknowledge packet receipt (e.g., UDP).
For them to be used in fault localization, they must be extended with simple ACK packets or some other type of coordination to detect when traffic is dropped; the ACKs do not need to be used for any other purpose.
The rest of this paper assumes TCP traffic.

End hosts attribute the drops to paths as described in the preceding section.
To handle uncertainty during the mid-update period, they evenly attribute drops to all applicable predictions.
For example, suppose that a single TCP connection has 100 packets.
If there are two possible versions, we attribute 50 packets to each path.
If there are four, we attribute 25.

\vspace{-1em}
\subsubsection{Implicating Components}

\nameFL uses path drop statistics to then implicate faulty components.
This step can be viewed as a classic inference problem: given observed drops, we infer drop rate for each component and then flag them as faulty if the rate is high.

\begin{figure}[t]
\centering
\epsfig{file=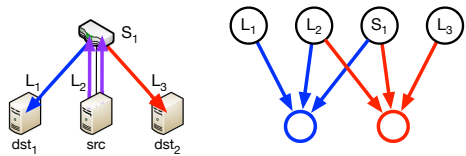, width=0.8\columnwidth}
\vspace{-1em}
\caption{Two flows (denoted by blue and red arrows) in a simple topology and how \name views the interactions between the flow and the underlying components.  Both flows start at the middle server and traverse $L_2$.  The blue flow takes $L_1$ while the red flow takes $L_2$.  From the interactions, if only the blue flow sees abnormal drops, $L_1$ is probably faulty.  Likewise for the red flow and $L_3$.}
\label{fig:system_model}
\end{figure}

\heading{System model.}
We model the impact of failures with a directed bipartite graph (\figref{system_model}).
The top partition consists of various network components.
We focus on links, switches, and routing table entries (RTEs), which are among the most common network failure granularities.
The bottom partition consists of paths and their drop statistics.
A component has an edge to a path if the path contains the component.

In this model, each path $i$ is observed to have transmitted $T_i$ and dropped ${D_i}$ packets, and each component $j$ has an unknown loss rate $x_j$ we want to infer.
We assume drops are independent for simplicity.
Our goal is to find loss rate for all components such that the sum of mis-predicted drops over all paths $L(\vec{x})$ is minimized.
\begin{align}
L(\vec{x})= \displaystyle\sum_{i\in\text{ all paths}} \left| 
D_i -   T_i \displaystyle\sum_{\substack{j\in Path(i)}} x_j
\right|
\end{align}

This formulation results in an estimated loss rate for every component, $\hat{x}_j$, rather than a binary up/down determination.
If any loss rate is above an operator-specified \textit{threshold}, it is flagged as a possible failure.
Though our model is relatively simple, it can be extended to handle additional component types or failure patterns through the same inference framework.

\heading{Localization Algorithm.}
\label{sec:localization_algorithm}
Finding component loss rates $\vec{x}$ that minimize $L(\vec{x})$ is a multivariate optimization problem.
We solve it using ideas from coordinate descent~\cite{coordinatedescent}.
Specifically, we initialize all components with zero loss rate, and greedily find the component loss rate $x_j$ that minimizes $L$ in isolation.
This process is iterative.

There are a few properties that our algorithm must satisfy in order to be practical.
First, it must handle the fact that retransmits can be caused by drops on both the forward \textit{and reverse} path, with no reliable way to differentiate between the two.
This is complicated by the fact that cumulative ACKs mean that drops on the reverse path are less likely to cause retransmissions than drops on the forward path.
Second, even if ACKs were accurate, drops may occur due to congestion and attribution can be inaccurate.
Finally, the algorithm must be able to compute the component loss rates very efficiently.

To address the difference between the forward and reverse path, we consider the two halves separately.
Note that the reverse path can be predicted from headers by swapping the source and destination addresses/ports.
When calculating the optimal drop rate for a particular component, we conservatively consider only forward paths as congestion statistics on them are much more accurate.
However, after we greedily choose the component that minimizes $L$,
its drop rate can be used to explain drops of flows that cross it in either direction.
Assuming sufficiently diverse traffic, all paths should be covered by some connection's forward path.

The other challenges are handled by the procedure:
\begin{vinenum}
\item Initialize $S$ to be the set of all components, and set the drop rate $x_j=0,\forall j\in S$.
\item For each $j$ in $S$, consider all forward paths it touches, $i\in \textit{FwdPaths}(j)$.  Find the loss rate for the component, $\hat{x}_j$, that minimizes $L$ assuming all other $x_k$ are unchanged:
\begin{align}
\hat{x}_j = \argmin_{x}\sum_{i} \left|D_{i} - T_{i} \left(x + \sum_{k \not= j} x_k \right)\right|
\end{align}
Note that computing this step is very efficient.
Because $\sum_i |D_i - T_i x|$ is piecewise linear,
we only need to check values of $x$ that make one of those terms inside the summation $|D_i - T_i x|=0$.
Further, the function is convex, implying that a binary search can find the global minimum.
\item Given the candidate $\hat{x}_j$ for all $j\in S$, pick the component $j'$ that minimizes $L$ and fix its drop rate $x_{j'}=\hat{x}_{j'}$.
\item Remove all explained drops from $\textit{FwdPaths}(j') \cup \textit{RvsPaths}(j')$ and remove $j'$ from $S$.
\item If some paths have unexplained drops above \textit{threshold} and max iteration not yet reached, repeat from step 2.
\end{vinenum}

\subsection{\nameLB: Load Balancing}
\label{sec:loadbalancing}

The second application we explore, \nameLB, demonstrates the emulation of in-network load balancing (specifically, CONGA~\cite{conga}) in a \name architecture.
We note that dynamic load balancing is a well-studied area with many other proposals implemented at both the end host and in the network.
The choice of protocols is therefore not an endorsement, but instead an opportunity to study the \name-friendly emulation of a routing algorithm with ``frequently changing inputs" as defined in \secref{design_other}.

At a high level, CONGA switches perform two functions.
First, they tag passing packets with congestion metrics and feed that information back to the source ToR.
Second, the source ToR waits for a sufficiently long inter-packet gap, rerouting each \textit{flowlet} toward the least-congested path.
In \nameLB, we separate these two functions explicitly and offload the second (flowlet rerouting) to end hosts.

\subsubsection{Collecting Congestion Metrics}

Like CONGA, \nameLB gathers congestion metrics via in-band feedback.
As a packet travels from the source to the destination ToR, switches tag the packet with their current load if it is larger than previously tagged values (see \cite{conga} for details).
The destination ToR then feeds these path-level congestion metrics back to the source ToR by piggybacking the information on normal traffic.
For every feedback-carrying packet, the destination ToR sends a single path-level metric, choosing amongst them in round-robin fashion.

At the end of the above process, the source ToRs have a lowest-utilized path toward every destination (multiple in the case of ties).
In addition, as none of these operations affect routing, they can all be done without losing predictability.

Where we begin to differ from CONGA is with an extra step to transfer the congestion metrics to servers in the source rack.
\nameLB uses two mechanisms.
First, the ToR switch uses incoming traffic to the rack to opportunistically piggyback the congestion metrics to its member servers.
For every packet sent to a member server, the ToR switch tags it in its egress pipeline with a (destination rack, best path to the rack) tuple.
The destination rack is chosen in a round-robin fashion, and if there are multiple best paths, a hash of the packet header is used to break tie.
In theory, congestion metrics kept at servers would be less up-to-date compared to what their ToR switches maintain.
However, for servers that communicate often with others, their congestion metrics would be refreshed timely by incoming ACKs or data packets.
The second mechanism allows servers to query their ToR for the best path to a destination leaf using UDP packets.
Servers send those requests to ToRs at connection setup in parallel with their SYN packets.
The on-demand query allows servers to steer to good default paths after long silence.

\vspace{-1em}
\subsubsection{Flowlet Steering}

In parallel with congestion metric collection, end hosts in \nameLB monitor inter-packet spacing to detect flowlets~\cite{LetFlow}.
For every new flowlet, the server steers the flowlet toward the destination's last `best path'.
Since end hosts know when and where flowlets are rerouted, predictability is maintained.
Extension of \nameFL to flowlets instead of flows is straightforward.

Our approach maintains the metrics and features of CONGA with minimal extra overhead (some additional header data on ToR-server packets).
Pushing the decision to servers increases the latency of feedback and decreases the rate at which feedback arrives at the decision point, but per our evaluation in \secref{repair_eval}, the effects are negligible.

\vspace{-1em}
\section{Evaluation}
\vspace{-1em}

We leverage a few evaluation platforms.
To evaluate the feasibility of predictable networks we implement one on top of a large data center at Facebook.
To test the performance of failure localization in a more controllable environment, we use an 80-machine Cloudlab testbed.
Finally, to test the relative performance of \nameLB and CONGA, we simulate the necessary hardware changes in ns-3.
We show that:

\begin{vinlist}
\item Some of today's networks are already predictable without modifying hardware or nonparticipating end hosts.
\item \nameFL is effective in locating a diverse set of failures and is robust to topology updates. %
\item \nameLB can closely approximate 
the performance of state-of-the-art in-network approaches.
\end{vinlist}
\vspace{-1em}

\subsection{Feasibility of \name}
\label{sec:prototype}

We begin by demonstrating the feasibility of prediction using a prototype implementation of \name on a large production data center at Facebook.
The data center has upwards of one hundred thousand devices and hosts a variety of applications, from frontend web servers and caching to backend storage and data analytics~\cite{fb_datacenter}.
For the most part, servers are connected into the network with a single 10\,Gbps link, while interconnect switches use 40\,Gbps links.

All of the switches are based on chipsets from one of the biggest manufacturers of switch ASICs, but span multiple vendors.
These switches support a diverse set of configurations for routing.
Just for ECMP, the options included flexible field selection, hash seeds, pre- and post-processing steps, and many possible hash functions.
Our predictor faithfully reproduced the path computation pipeline of these switches along with the effects of all of these configuration options.
It gathered the options from switches in order to perform predictions.

Our prototype did not require any modifications to either switch configurations or OS configurations---the network, as configured, already approximated a predictable network.
We also verified the feasibility of our system on top of a testbed of Cavium switches and ASICs, but we omit those results here due to space constraint.

\subsubsection{Predicting Paths}

To test the accuracy of our predictor, we ran UDP traceroutes between servers in the data center, and compared ground truths with our path prediction engine.
The ToR switch was already configured with an ECMP group.
When replicating the relevant configuration options within our path
deduction engine, we are able to replicate 100\% of the results
recorded by the UDP traceroute experiment.
We also built the predictor's inverse for the purpose of efficiently generating headers for a target path.

\begin{figure}
\centering
\epsfig{file=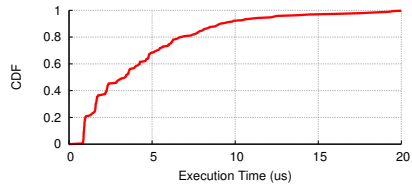, width=0.7\columnwidth}
\caption{CDF of the time it takes to compute a header for a specific path.  The topology we use is a fully-deployed version of a recently published data center architecture~\cite{fb_datacenter}.}
\label{fig:prediction_overhead}
\end{figure}

\heading{Overhead of prediction.}
In addition to verifying that our prediction engine can accurately predict paths, we also tested the efficiency of the engine when trying to find a header for a particular network path.
For this experiment, we chose a fixed source and destination server in the data center.
There were hundreds of potential paths between the two machines through a topology similar to the one described in~\cite{fb_datacenter}.
As our inverse function is only able to reverse a single switch's routing function at a time, generating a header for a specific multi-hop path involved a multi-step process.
The first step is to use the inverse function to generate a valid header for one of the switches (with a preference for the switch with the largest ECMP group).
We then use our predictor to check the generated header on the second switch.
If the header maps to the correct routing choice and does not require a reserved port, we accept the header, otherwise, we start again.

\figref{prediction_overhead} shows a CDF of the execution time of the above algorithm on a server with a 2.60\,GHz Intel Xeon E5-2670.
For the test, we picked a specific path and gave the generator control over the UDP source port of the header
We then tracked the time it took to find a target header for the given path.
The algorithm was always able to find a valid header with a median execution time of 3.4\,$\upmu$s.

\vspace{-1em}
\subsubsection{Controlling Paths}

To demonstrate path control, we implemented and deployed to a small set of nodes in the aforementioned data center an \texttt{iptables} user-space application called `ECMP-interpose'.
ECMP-interpose automatically and transparently modifies connection parameters when a TCP timeout occurs.
Modifications are as described in \secref{endhosts}.

More concretely, ECMP-interpose installs rules into \texttt{iptables}
that match on relevant incoming and outgoing TCP traffic and relay
packets via the NFQUEUE target to a user-level packet queue.
For each connection, we install rules matching the
TCP source port into the INPUT and OUTPUT chains in the filter table
on both end hosts as described previously.
\texttt{iptables} allows rules for several connections to be consolidated via range and set matches for performance.
After modification, it computes the new TCP checksum and relays
packets back to the kernel.

\begin{figure}
\centering
\epsfig{file=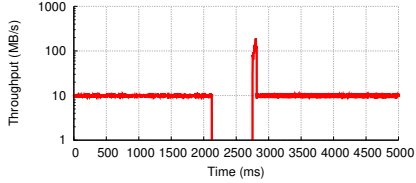, width=0.7\columnwidth}
\caption{Throughput of a constant-rate connection in the presence of a failure.  Upon TCP timeout, the sender transparently switches paths. The post-failure spike was due to a rush of ACKs.}
\label{fig:failover_exp}
\end{figure}

\heading{Effect of rerouting.}
We evaluate our prototype with a simple rerouting experiment.
We run a constant-rate TCP connection from a source to a destination in a different cluster in the same data center.
At $\sim$2 seconds, we fail the connection.
When the sender gets a timeout (via \texttt{tcp\_retransmit\_timer()} in the Linux networking stack), ECMP-interpose automatically switches to an alternate path.
While the timeout took $\sim$500\,ms, ECMP-interpose's switchover was nearly instantaneous---a more aggressive failure detection method would have minimized interruption of connectivity.
We conclude that we can successfully and transparently and selectively change ECMP routes of live connections by interposing on these connections and modifying their port numbers.
Route changes are instant and stable.

\vspace{-1em}
\subsection{\nameFL Evaluation}
\label{sec:localize_eval}

We evaluate \nameFL by asking several questions:

\begin{vinlist}
\item Can we localize different types of failures and how sensitive is our approach to the failure's drop rate?
\item How much does the aggregation period length matter?
\item What about multiple, potentially heterogeneous faults?
\item How much does a stale view of topology affect results?
\end{vinlist}

\heading{Testbed.}
We answer these questions using an 80-machine Cloudlab~\cite{cloudlab} testbed.
The machines were interconnected via a 10\,Gbps network.
Each physical machine emulates either a server or a predictable software switch. %
We use GRE-tunnel~\cite{gre_tunnel} to implement an overlay Clos network.
The resulting topology has 12 racks with 4 servers each.
The racks' 12 ToRs are split into 3
clusters with 4 aggregation switches in each.
Each aggregation switch connects to two core switches, for a total of 8 core switches.
We use Linux tc to limit link bandwidth to 1\,Gbps, 
and emulate RED queues with ECN marking threshold at 30\,KB~\cite{DCTCP}.
We configure Linux to use DCTCP.

\begin{figure}[t!]
\centering
\epsfig{file=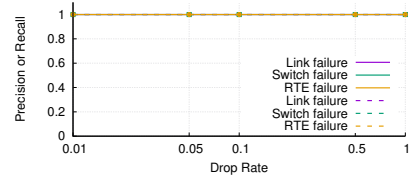, width=0.7\columnwidth}
\caption{Precision and recall for a single failure and various drop rates from 1\% to 100\%.  We considered link failures, switch failures, and failures of individual routing table entries over a window of 10\,s.}
\label{fig:localize_single}
\end{figure}

To collect drop statistics, we use Linux eBPF~\cite{eBPF} with \texttt{bcc}~\cite{bcc}.
Unless stated otherwise, drop statistics were polled every 10 seconds.
We use recall and precision, averaged over 50 runs, as metric for fault localization.
Recall is the percentage of faults that have been predicted
and precision is the percentage of predictions that are correct.

\heading{Workload.}
We generate traffic according to a realistic workload based on empirically observed traffic patterns in deployed data centers~\cite{DCTCP}.
The web search workload is heavy-tailed: a small fraction of flows contribute most of the traffic.
Flows arrive according to a Poisson process between server pairs evenly.
We inject an offered load of $40\%$ of total host access link bandwidth.

\heading{Failures.}
We injected failures into the network at random time
while running our failure localization application in the background.
The set of failures we tested were drawn from those emphasized by recent literature~\cite{F10,Everflow,NetPilot} and they cover the range of failure behaviors listed in \secref{case_endhost}.
In particular, several types of components can fail silently in our testbed: links, switches, and individual routing table entries.
Failures can either be fail-stop or stochastic with some drop rate.

\subsubsection{Localization of a Single Failure}
\label{sec:localize_single}

We first evaluate our localization precision and recall for a single failure.
We inject failures at either a link, switch, or routing table entry, at random location.
We tested various drop rates ranging from 1\% to 100\%.
The mean time from failure to end host notification was $21.8$ seconds (much of this was due to our use of a 10\,s aggregation period).

\figref{localize_single} shows that, for most cases, our algorithm has perfect precision and recall.
This is because greedy is optimal when there is only a single instance of failure.

\subsubsection{Effect of Aggregation Period Length}
\label{sec:eval_window}

\begin{figure}[t!]
\centering
\epsfig{file=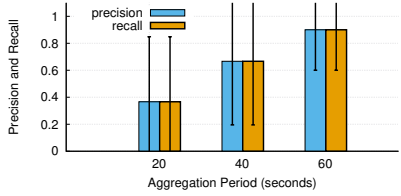, width=0.7\columnwidth}
\caption{Precision and recall for detecting a 0.1\% failure over a range of aggregation periods.}
\label{fig:localize_speed}
\end{figure}

Lower-rate failures can be detected by increasing the aggregation period.
This represents a tradeoff.
A longer aggregation period can filter out transient noise from the data, making our predictions more accurate, and also decrease the overhead of collection.
This, however, increases failure detection latency.

Up until now, we have been using a 10 seconds aggregation period.
In this experiment, we test how long the aggregation period needs to be to detect a 0.1\% loss rate link failure.
In principle, any persistent failure with loss rate greater than the steady-state congestion loss rate of the network can be located with a long enough aggregation period.

In \figref{localize_speed}, we show the precision and recall for windows ranging from 20\,s to 60\,s.
As we aggregate over longer period, detection becomes more accurate, reaching 90\% precision and recall for 0.1\% loss rate with a 60 second aggregation period.

\subsubsection{Multiple, Simultaneous Failures}
\label{sec:localize_many}

\nameFL also extends to multiple simultaneous, possibly heterogeneous, failures.
We injected a random mix of failures and look at the precision and recall for our algorithm.
The failures are randomly chosen: they can be link failures, switch failures, or routing table corruptions.
Their drop rates are sampled uniformly between 1\% and 100\%.

\begin{figure}[t!]
\centering
\epsfig{file=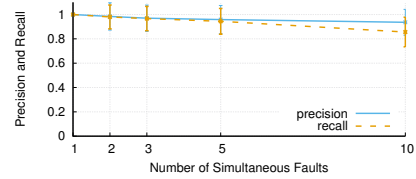, width=0.7\columnwidth}
\caption{Average precision and recall for simultaneous failures in our testbed.  The failures are of a random type and rate.}
\label{fig:localize_many}
\end{figure}

\figref{localize_many} shows the average precision/recall for up to 10 simultaneous failures.
Across the experiments, our system maintains a precision above 0.95 and a recall above 0.85, even when failure count is high.
As the failure count increases, recall decreases.
This is expected as our algorithm is greedy and assumes that a few larger failures are more likely than many smaller failures.

\vspace{-1em}
\subsubsection{Impact of Stale State}
\label{sec:eval_mispredicts}

\begin{figure}[t!]
\centering
\epsfig{file=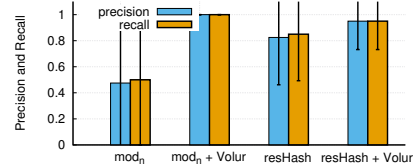, width=0.7\columnwidth}
\caption{The average precision and recall for detecting a 10\% switch failure in the presence of a topology change. With \name state dissemination, precision and recall increase to above 0.95 regardless of failover scheme ($\mathrm{mod}_\mathrm{n}$ or resHash).}
\label{fig:localize_topochange}
\end{figure}

Part of \name's design is that switches can make routing changes on-the-fly, as long as those changes are infrequent.
In this subsection, we evaluate the impact of stale state on failure localization.
More specifically, we try to locate a single 10\% drop rate failure in the presence of a topology-changing switch reboot.

We first configure a random aggregation or core switch %
to silently drop 10\% of packets.
Later, we reboot a random aggregation switch, which is properly detected/disseminated via the \name service.

We evaluated two different switch failover policies with and without state dissemination.
The first policy, `$\mathrm{mod}_\mathrm{n}$', remaps all flows using a simple modulo function.
The result is that most flows change paths after a failure.
The second, resilient hashing (`resHash'), uses a simple, predictable function that limits the number of flows that need to change paths after a single failure.

\figref{localize_topochange} shows that without resilient hashing or topology dissemination, precision and recall falls to around 0.5, with successes limited to cases where the failures are in separate subtrees and therefore most traffic is predicted correctly.
With resilient hashing, both numbers rise to 0.84 as resilient hashing avoids remapping every flow.
Thus, a large number of path predictions are still correct even with stale network state.
For both failover strategies, adding topology dissemination brings precision and recall back above 0.95.

\vspace{-1em}
\subsection{\nameLB Evaluation}
\label{sec:repair_eval}

We evaluate the performance of \nameLB with a 12-switch, 72-host ns-3 simulation.
We show that \nameLB achieves an average flow completion time (FCT) within 1.05x of CONGA at low to moderate load, and within 1.1x at very high load for both symmetric and asymmetric topologies.

\heading{Architecture.}
We used a 6-leaf 6-spine topology with 10\,Gbps links and 2:1 leaf oversubscription ratio.
In the \textit{symmetric} topology, all links have 10\,Gbps capacity.
In the \textit{asymmetric} topology, each leaf has 2 randomly picked uplinks out of 6 uplinks with half capacity.
In the worst case, a leaf to leaf path can have 4 paths out of 6 paths with only 5\,Gbps capacity.
The degree of asymmetry is high.

End hosts use DCTCP and queues use RED with ECN marking, with a threshold of 65\,MTU and 700\,KB (467\,MTU) buffers~\cite{DCTCP}.

\heading{Workload.}
We generated flows according to the enterprise workload in \cite{conga,LetFlow} with arrival rate to match different offered traffic load.
Traffic were generated using a simple client-server program at each host.
All traffic went through the spine to stress the load balancing properties of the fabric.
Each client established 6 persistent TCP connections with every server.

\heading{Methodology.}
We use flow completion time (FCT)  as evaluation metric.
We average over 5 runs.
We compare:
\begin{vinlist}
\item \textit{ECMP:} Our baseline is ECMP, in which each switch makes local, uniform-random load balancing decisions.
\item \textit{CONGA:} We use the default parameters: $Q=3$, $\tau=160$\,$\upmu$s, and flowlet timeout of $500$\,$\upmu$s.
We validated our implementation with the testbed results from~\cite{conga}.
\item \textit{\nameLB:} Finally, we implement \nameLB as described in \secref{loadbalancing}.  Where applicable, we use the same configuration parameters as our CONGA implementation.
\end{vinlist}

\begin{figure}[t!]
\centering
\subfloat[Avg. FCT on symm topology]{
\epsfig{file=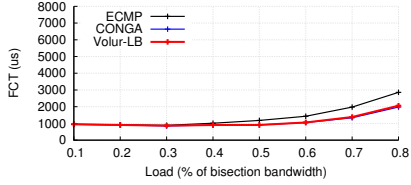, width=0.7\columnwidth}
\label{fig:fct_symm}
}

\vspace{-1.5em}
\subfloat[Avg. FCT on asymm topology with 1/3 half-capacity uplinks]{
\epsfig{file=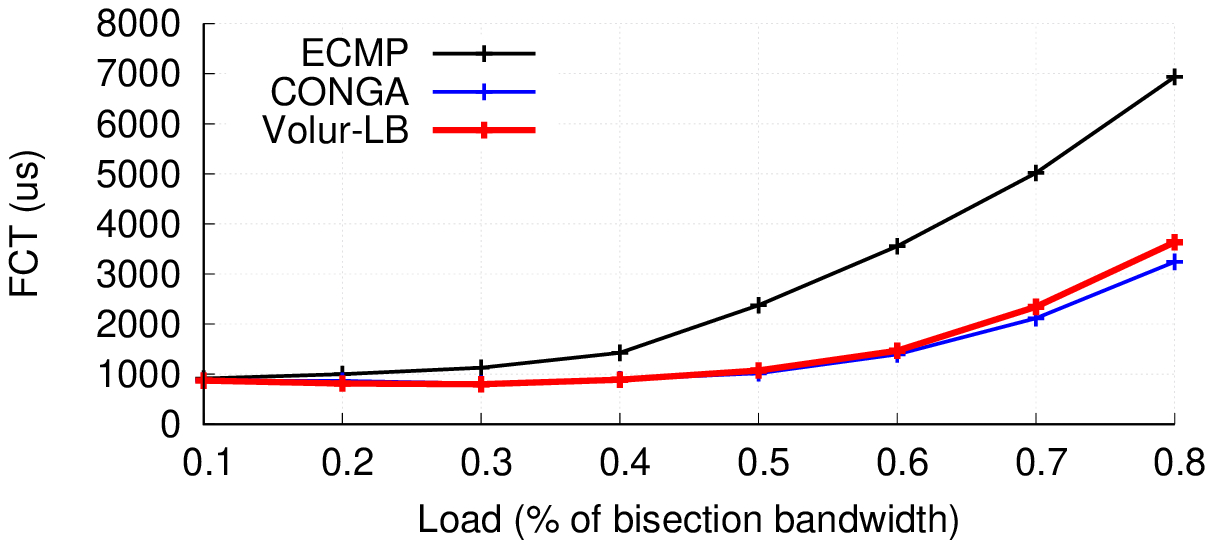, width=0.7\columnwidth}
\label{fig:fct_asymm}
}
\caption{\nameLB achieves almost identical FCT as CONGA for symm topology and within 1.1x for asymm topology.}
\label{fig:fct_lb}
\end{figure}

\topheading{Results.}
\figref{fct_symm} shows that \nameLB achieves almost identical average FCT to CONGA on symmetric topology for all load.
\figref{fct_asymm} shows that \nameLB achieves FCT within 1.1x of CONGA on asymmetric topology even for load as high as 80\%.

\vspace{-1em}
\section{Related Work}
\label{sec:related}
\vspace{-0.75em}

Over the years, the research community has pointed out deficiencies in data center network routing, both in load balancing and fault tolerance. %
Broadly speaking, proposed solutions fall into one of three categories:

\heading{Network control.}
The network is a straightforward place to address these deficiencies.
For load balancing, systems like CONGA~\cite{conga} and DRILL~\cite{drill} add functionality to switches so that they can react to traffic bursts at very short time scale.
For failure localization, a similar trend has been to augment in-network monitoring~\cite{SNMP, sFlow, AdaptiveNetFlow, PacketHistory, Everflow, FlowRadar}
Although these approaches are powerful, when the mechanisms themselves fail, end hosts are left with no recourse.

\heading{End host control.}
Another class of prior work attempts to address deficiencies at the end host.
A few of these propose to either work around the network's opacity or move bits of routing functionality to the end hosts.
In particular, LetFlow~\cite{LetFlow} and CLOVE~\cite{CLOVE} both make a case for end host load balancing as do several other approaches~\cite{flowbender,presto,ResilientLB}.
The same is true of failure detection~\cite{NetNORAD,Pingmesh}.
Our work is complementary to these systems as we seek to support future routing innovation so that new proposals are not hamstrung by ECMP's interface.

More extreme are proposals like XPATH~\cite{XPATH}, which argue for source routing in the data center.
These give full route control and visibility to the end hosts, but they come at the cost of essential features like fast in-network failover.

\heading{Network and end host coordination.}
The idea of an intelligent network assisting intelligent hosts has been explored in other areas.
For example, ECN-based transport~\cite{DCTCP,Floyd1993RandomED} provides end hosts with information about the utilization of the network.

In a similar vein, other proposals have sought to increase visibility into the network by tagging packets with their path as they pass switches~\cite{roy17passive,pathdump}.
The caveat with this approach is that locating failures with tags requires successful delivery;
if none of the target packets make it through to the destination, the route will remain hidden.

In comparison, \name provides end hosts with a complete and up-to-date view of routing in the network, greatly expanding the options for and efficacy of end host functions.

\tr{
\vinpresection
\section{Discussion}
\vinpostsection
\label{sec:extensions}

For simplicity, we have thus far focused on typical Clos topologies and simple switch configurations.
Real data centers come in many shapes and sizes---too many to cover in detail.
In this section, we point out a few interesting variations.

\vinpresubsection
\subsection{Redundant downward paths}
\vinpostsubsection

One way to increase network fault tolerance at the cost of scalability is to add redundant links between root switches and their subtrees.
\figref{XXX} shows an example topology from the VL2 paper that exhibits this property.
In the figure, fabric switches connect to more than one cluster switch in each subtree.
This removes a fundamental weakness in traditional Clos networks: that there is only a single path from every root to every leaf.

In \name, this redundancy implies that we must be able to steer traffic in the downward direction---something that the design of the previous section cannot accomplish because with the Trident 2, switches use the same hash keys and hash functions on the upward direction and the downward direction.
This clashes with our goal of allowing independent control over every ECMP decision in the network.

To enable switches to hash two different packets differently, we use encapsulation.
The original header is used to control the downward path while the outer header is used to control the upward path.
The last switch on the upward path is responsible for decapsulating the packet.

An easy way to implement this would be to use an MPLS header.
Today's switches already have the ability to hash on and decapsulate MPLS labels.
Sources would simply encapsulate the original packet with an MPLS label that resulted in the correct ECMP decisions on all switches on the upward path.

More generally, encapsulation allows us to stitch a theoretically arbitrary number of path segments together.
See \secref{related} for a brief discussion of the limits of deep MPLS encapsulation.
\vlcomment{TODO: we can do this with flexible hashing}

\tr{
\vinpresubsection
\subsection{Arbitrary Topologies}
\vinpostsubsection

Unstructured topologies are not amenable to any clean division of hash groups.
Unfortunately, it is these topologies that need \name the most---routing table scalability is one of the primary challenges in these types of networks.
For these topologies, we introduce two tools:

\begin{vinlist}
\item \emph{Encapsulation:} The mechanism described in the previous section works as long as no paths require more than one ECMP decision in any given hash group.
We henceforth refer to paths that satisfy this ``one decision per group'' property as path segments.  Encapsulation allows us to stitch path segments together.
\item \emph{Offline Checking:} For random or otherwise unstructured topologies, division of switches into hash groups is exceedingly difficult.
For these random networks, we propose to do random assignment of hash groups with a offline check to verify the quality of the random assignment.
\end{vinlist}

We therefore adapt \name to random topologies by first randomly assign bit ranges and hash seeds to switches.
Assuming sufficient randomness, a good hash function and a large enough set of usable header bits (i.e., address space), many useful between a given source and destination should be addressable.

To verify that the hash settings indeed provide good connectivity without encapsulation, we add a verification step.
When designing the topology offline, we manually check the shortest $N$ edge-disjoint paths between every source and every destination ToR.
These paths will provide a few good load-balancing options.
Verification of a path involves solving a system of $n$ linear equations, where $n$ is the number of switches on the path (which should be small given that we are looking for shortest paths).
The complexity of the entire operation is then $O(Nr^2n^3)$, where $r$ is the number of ToR switches and we use simple Gaussian elimination to solve the system of equations.
Note again, that this is done entirely offline and only on physical expansion of the network.

Failures can cause some of these paths to become unusable and can even cause otherwise correct paths to become unroutable due to changes in the modulo divisor.
We handle this by dynamically choosing the next shortest path and/or using encapsulation to route along previously unroutable paths.
}

\vinpresubsection
\subsection{Resilient hashing}
\vinpostsubsection
\label{sec:resilienthashing}

In a traditional switch, when a failure occurs, the number of next hop entries in the routing table may change, causing the hashing function (or more specifically, the modulus step at the end) to change as well.
This in turn causes widespread rebinding of almost all flows---even those on ports that are unaffected by the failure.

To avoid rebinding, some recent switches have implemented a feature called resilient hashing, which attempts to avoid as much rebinding as possible.
It does so by leveraging a separate `flow table'.
When a new flow arrives, the hashing mechanism will create an entry in the flow table that maps the flow to a physical egress port.
Whenever a packet arrives, the switch will check the flow table before recomputing the hash value.
Thus, when a failure occurs, the switch will continue to use the cached entry for all flows that are unaffected by the failure.
Flows on failed links are rebound to the surviving links.
Similarly, when a new link is activated, resilient hashing will attempt to minimize the number of rebound flows.
Specifically, it will choose a number of existing flows to rebind, but not touch any other flows.

In the context of \name, this feature is both a positive and a negative.
On one hand, it means that topology changes cause significantly less path uncertainty.
On the other, some flows that are not even on the changed path may be moved permanently.
This is another instance where further cooperation from switch ASIC manufacturers would be helpful.
If we know the algorithm for rebinding, it could be included in the path deduction and control algorithms.
The other alternative is to disable this feature.
 }

\vspace{-1.25em}
\section{Conclusion}
\vspace{-1em}

This paper presents an architecture that facilitates the co-existence of route control both in the network and at end hosts.
Our results show that the architecture is both feasible and flexible.
Using it, we demonstrate a failure handling mechanism that is both accurate and responsive.
We also demonstrate an end host load balancing mechanism that emulates state of the art in-network approaches predictably.
Finally, we verify the feasibility of our approach by building a test deployment on a large, otherwise unmodified production data center network.

{
\small
\bibliographystyle{abbrv}
\bibliography{ref} 

\begin{thebibliography}{10}

\bibitem{NetNORAD}
A.~Adams, P.~Lapukhov, and H.~Zeng.
\newblock {NetNORAD}: Troubleshooting networks via end-to-end probing, 2016.
\newblock
  \url{https://code.facebook.com/posts/1534350660228025/netnorad-troubleshooting-networks-via-end-to-end-probing/}.

\bibitem{conga}
M.~Alizadeh, T.~Edsall, S.~Dharmapurikar, R.~Vaidyanathan, K.~Chu,
  A.~Fingerhut, V.~T. Lam, F.~Matus, R.~Pan, N.~Yadav, and G.~Varghese.
\newblock {CONGA}: Distributed congestion-aware load balancing for datacenters.
\newblock In {\em SIGCOMM}, 2014.

\bibitem{DCTCP}
M.~Alizadeh, A.~Greenberg, D.~A. Maltz, J.~Padhye, P.~Patel, B.~Prabhakar,
  S.~Sengupta, and M.~Sridharan.
\newblock Data center {TCP} ({DCTCP}).
\newblock In {\em SIGCOMM}, 2010.

\bibitem{fb_datacenter}
A.~Andreyev.
\newblock Introducing data center fabric, the next-generation {Facebook} data
  center network.
\newblock \url{https://code.facebook.com}, Nov. 2014.

\bibitem{P4}
P.~Bosshart, D.~Daly, G.~Gibb, M.~Izzard, N.~McKeown, J.~Rexford,
  C.~Schlesinger, D.~Talayco, A.~Vahdat, G.~Varghese, and D.~Walker.
\newblock P4: Programming protocol-independent packet processors.
\newblock {\em SIGCOMM CCR}, 44(3):87--95, July 2014.

\bibitem{CumulusResilient}
P.~Bratach and P.~Lumbis.
\newblock {Equal Cost Multipath Load Sharing - Hardware ECMP}, 2017.
\newblock
  \url{https://docs.cumulusnetworks.com/display/DOCS/Equal+Cost+Multipath+Load+Sharing+-+Hardware+ECMP}.

\bibitem{SNMP}
J.~Case, R.~Mundy, D.~Partain, and B.~Stewart.
\newblock Introduction and applicability statements for internet standard
  management framework, 2002.
\newblock \url{https://tools.ietf.org/html/rfc3410}.

\bibitem{eBPF}
J.~Corbet.
\newblock Extending extended {BPF}, 2014.
\newblock \url{https://lwn.net/Articles/603983/}.

\bibitem{davies2010traffic}
M.~Davies.
\newblock Traffic distribution techniques utilizing initial and scrambled hash
  values, Oct.~26 2010.
\newblock US Patent 7,821,925.

\bibitem{AdaptiveNetFlow}
C.~Estan, K.~Keys, D.~Moore, and G.~Varghese.
\newblock Building a better netflow.
\newblock In {\em SIGCOMM}, 2004.

\bibitem{Floyd1993RandomED}
S.~Floyd and V.~Jacobson.
\newblock Random early detection gateways for congestion avoidance.
\newblock {\em IEEE/ACM Trans. Netw.}, 1:397--413, 1993.

\bibitem{drill}
S.~Ghorbani, Z.~Yang, B.~Godfrey, Y.~Ganjali, and A.~Firoozshahian.
\newblock Drill: Micro load balancing for low-latency data center networks.
\newblock In {\em SIGCOMM}, 2017.

\bibitem{bcc}
B.~Gregg.
\newblock {BCC}: Dynamic tracing tools for linux, 2017.
\newblock \url{https://iovisor.github.io/bcc/}.

\bibitem{Pingmesh}
C.~Guo, L.~Yuan, D.~Xiang, Y.~Dang, R.~Huang, D.~Maltz, Z.~Liu, V.~Wang,
  B.~Pang, H.~Chen, Z.-W. Lin, and V.~Kurien.
\newblock Pingmesh: A large-scale system for data center network latency
  measurement and analysis.
\newblock In {\em SIGCOMM}, 2015.

\bibitem{PacketHistory}
N.~Handigol, B.~Heller, V.~Jeyakumar, D.~Mazi\`{e}res, and N.~McKeown.
\newblock I know what your packet did last hop: Using packet histories to
  troubleshoot networks.
\newblock In {\em NSDI}, 2014.

\bibitem{presto}
K.~He, E.~Rozner, K.~Agarwal, W.~Felter, J.~Carter, and A.~Akella.
\newblock Presto: Edge-based load balancing for fast datacenter networks.
\newblock In {\em SIGCOMM}, 2015.

\bibitem{SWAN}
C.-Y. Hong, S.~Kandula, R.~Mahajan, M.~Zhang, V.~Gill, M.~Nanduri, and
  R.~Wattenhofer.
\newblock Achieving high utilization with software-driven wan.
\newblock In {\em SIGCOMM}, 2013.

\bibitem{rfc2992}
C.~Hopps.
\newblock Analysis of an equal-cost multi-path algorithm.
\newblock RFC 2992 (Informational), 2000.

\bibitem{XPATH}
S.~Hu, K.~Chen, H.~Wu, W.~Bai, C.~Lan, H.~Wang, H.~Zhao, and C.~Guo.
\newblock Explicit path control in commodity data centers: Design and
  applications.
\newblock In {\em NSDI}, 2015.

\bibitem{gre_tunnel}
B.~Hubert.
\newblock Gre and other tunnels, 2017.
\newblock \url{http://lartc.org/howto/lartc.tunnel.gre.html}.

\bibitem{B4}
S.~Jain, A.~Kumar, S.~Mandal, J.~Ong, L.~Poutievski, A.~Singh, S.~Venkata,
  J.~Wanderer, J.~Zhou, M.~Zhu, J.~Zolla, U.~H\"{o}lzle, S.~Stuart, and
  A.~Vahdat.
\newblock {B4}: Experience with a globally-deployed software defined wan.
\newblock In {\em SIGCOMM}, 2013.

\bibitem{flowbender}
A.~Kabbani, B.~Vamanan, J.~Hasan, and F.~Duchene.
\newblock {FlowBender}: Flow-level adaptive routing for improved latency and
  throughput in datacenter networks.
\newblock In {\em CoNEXT}, 2014.

\bibitem{kalkunte2010high}
M.~Kalkunte.
\newblock High speed trunking in a network device, Mar.~16 2010.
\newblock US Patent 7,680,107.

\bibitem{CLOVE}
N.~Katta, M.~Hira, A.~Ghag, C.~Kim, I.~Keslassy, and J.~Rexford.
\newblock Clove: How i learned to stop worrying about the core and love the
  edge.
\newblock In {\em HotNets}. ACM, 2016.

\bibitem{rfc5880}
D.~Katz and D.~Ward.
\newblock Bidirectional forwarding detection {(BFD)}, 2010.
\newblock \url{https://tools.ietf.org/html/rfc5880}.

\bibitem{AWSgray}
A.~L\^e-Qu\^oc.
\newblock Learning from {AWS'} gray failures.
\newblock \url{https://www.datadoghq.com/blog/gray-aws-failures/}, October
  2013.

\bibitem{lgorzata2004survey}
M.~{\l{}}gorzata Steinder and A.~S. Sethi.
\newblock A survey of fault localization techniques in computer networks.
\newblock {\em Science of computer programming}, 53(2):165--194, 2004.

\bibitem{FlowRadar}
Y.~Li, R.~Miao, C.~Kim, and M.~Yu.
\newblock Flowradar: A better netflow for data centers.
\newblock In {\em NSDI}, 2011.

\bibitem{F10}
V.~Liu, D.~Halperin, A.~Krishnamurthy, and T.~Anderson.
\newblock F10: A fault-tolerant engineered network, 2013.

\bibitem{DLB}
B.~Matthews, B.~Kwan, and P.~Agarwal.
\newblock Dynamic load balancing, Jan.~15 2013.
\newblock US Patent 8,355,328.

\bibitem{Openflow}
N.~McKeown, T.~Anderson, H.~Balakrishnan, G.~Parulkar, L.~Peterson, J.~Rexford,
  S.~Shenker, and J.~Turner.
\newblock Openflow: Enabling innovation in campus networks.
\newblock In {\em SIGCOMM}, 2008.

\bibitem{Gestalt}
R.~N. Mysore, R.~Mahajan, A.~Vahdat, and G.~Varghese.
\newblock Gestalt: Fast, unified fault localization for networked systems.
\newblock In {\em USENIX ATC}, pages 255--267, Philadelphia, PA, June 2014.
  USENIX Association.

\bibitem{sFlow}
P.~Phaal, S.~Panchen, and N.~McKee.
\newblock {InMon} corporation's {sFlow}: A method for monitoring traffic in
  switched and routed networks.
\newblock RFC 3176 (Informational), 2001.

\bibitem{cloudlab}
R.~Ricci, E.~Eide, and {The CloudLab Team}.
\newblock Introducing {CloudLab}: Scientific infrastructure for advancing cloud
  architectures and applications.
\newblock {\em {USENIX} {;login:}}, 39(6), Dec. 2014.

\bibitem{MPLS}
E.~Rosen, A.~Viswanathan, and R.~Callon.
\newblock Multiprotocol label switching architecture.
\newblock RFC 3031, 2001.

\bibitem{roy17passive}
A.~Roy, J.~Bagga, H.~Zeng, and A.~C. Snoeren.
\newblock Passive realtime datacenter fault detection and localization.
\newblock In {\em NSDI}, 2017.

\bibitem{Saltzer1981EndToEndAI}
J.~H. Saltzer, D.~P. Reed, and D.~D. Clark.
\newblock End-to-end arguments in system design.
\newblock {\em ACM Trans. Comput. Syst.}, 2:277--288, 1981.

\bibitem{LocalFlow}
S.~Sen, D.~Shue, S.~Ihm, and M.~J. Freedman.
\newblock Scalable, optimal flow routing in datacenters via local link
  balancing.
\newblock In {\em CoNEXT}, 2013.

\bibitem{pathdump}
P.~Tammana, R.~Agarwal, and M.~Lee.
\newblock Simplifying datacenter network debugging with pathdump.
\newblock In {\em OSDI}, 2016.

\bibitem{LetFlow}
E.~Vanini, R.~Pan, M.~Alizadeh, T.~Edsall, and P.~Taheri.
\newblock Let it flow: Resilient asymmetric load balancing with flowlet
  switching.
\newblock In {\em NSDI}, 2017.

\bibitem{coordinatedescent}
S.~J. Wright.
\newblock Coordinate descent algorithms.
\newblock {\em Mathematical Programming}, 151(1):3--34, 2015.

\bibitem{NetPilot}
X.~Wu, D.~Turner, G.~Chen, D.~Maltz, X.~Yang, L.~Yuan, and M.~Zhang.
\newblock {NetPilot}: Automating datacenter network failure mitigation.
\newblock In {\em SIGCOMM}, 2012.

\bibitem{ResilientLB}
H.~Zhang, J.~Zhang, W.~Bai, K.~Chen, and M.~Chowdhury.
\newblock Resilient datacenter load balancing in the wild.
\newblock In {\em To appear in SIGCOMM}, 2017.

\bibitem{WCMP}
J.~Zhou, M.~Tewari, M.~Zhu, A.~Kabbani, L.~Poutievski, A.~Singh, and A.~Vahdat.
\newblock {WCMP}: Weighted cost multipathing for improved fairness in data
  centers.
\newblock In {\em EuroSys}, 2014.

\bibitem{Everflow}
Y.~Zhu, N.~Kang, J.~Cao, A.~Greenberg, G.~Lu, R.~Mahajan, D.~Maltz, L.~Yuan,
  M.~Zhang, B.~Y. Zhao, and H.~Zheng.
\newblock Packet-level telemetry in large datacenter networks.
\newblock In {\em SIGCOMM}, 2015.

\end{thebibliography}
}

\end{document}